\documentclass[a4paper]{jpconf}
\usepackage{graphicx}

\newcommand{\mpt}{\mbox{$p_\mathrm{T}\,$}}
\newcommand{\kt}{\mbox{$k_\mathrm{T}\,$}}
\newcommand{\xe}{\mbox{$x_\mathrm{E}\,$}}

\newcommand{\piz}{\mbox{$\pi^0\,$}}
\newcommand{\iaa}{\mbox{$I_{\mathrm{AA}}\,$}}
\newcommand{\raa}{\mbox{$R_{\mathrm{AA}}\,$}}
\newcommand{\rg}{\mbox{$R_{\gamma}\,$}}

\begin{document}
\title{Jet Fragmentation in Vacuum and Medium with $\gamma$-hadron Correlations in PHENIX}

\author{Matthew Nguyen}

\address{Research Fellow, CERN, Geneva, Switzerland}

\ead{Matthew.Nguyen@cern.ch}

\begin{abstract}

Jet fragmentation in $p$+$p$ and Au+Au collisions is studied via back-to-back correlations of direct photons and charged hadrons.  The direct photon correlations are obtained by statical subtraction of the background from decay photons.  Results on the nuclear modification to the associated charged hadron yields are reviewed.   Further studies of jet fragmentation in $p$+$p$ using isolated direct photons are also presented.   A \kt-smeared LO pQCD calculation is used to interpret the data.   The sensitivity of the data to the underlying fragmentation function is tested and the results are found to be compatible with expectations of a sample dominated by quark jet fragmentation.  

\end{abstract}

\section{Introduction}
The suppression of high \mpt hadrons and the modification of back-to-back di-hadron correlations in nuclear collisions are generally understood to be manifestations of in-medium parton energy loss~\cite{ppg003,ppg032}.   The quantitive interpretation of such measurements is, however, complicated by the fact that the initial parton energy is unknown.   The dominant mechanism for direct photon production at LO in pQCD is the quark-gluon Compton scattering process, $q+g \rightarrow q+\gamma$.  Compared to partons, photons have a small cross section to interact with the medium.    Modulo higher order effects, the direct photon can hence be used to determine the initial \mpt of the recoil parton before energy loss, since they exactly balance.  Using the photon as a proxy for the parton, the fragmentation function of the recoil jet, which may be effectively modified by the medium, can be studied in detail~\cite{arleo}.

\section{Photon-Hadron Correlations}

  Experimentally, the challenge of direct photon measurements is to separate them from the large background of photons produced from hadron decays.  Direct photon correlations with charged hadrons may be obtained by a statistical subtraction of the decay component as follows.   First the per-trigger yield ($Y$) of all photons is measured, which is the number of photon-hadron pairs divided by the number of photons, i.e., $Y_{\rm total} \equiv N_{\gamma-h}/N_{\gamma}$.   This quantity can be decomposed into contributions from direct and decay sources:

\begin{equation}
Y_{\rm total} = \frac{N_{\rm direct}}{N_{\rm total}}Y_{\rm direct} + \frac{N_{\rm decay}}{N_{\rm total}}Y_{\rm decay},
\end{equation}

\noindent where direct is taken to signify any photon not from hadron decay.  We may solve for $Y_{\rm direct}$,

\begin{equation}
Y_{\rm direct} = \frac{\rg}{\rg-1} Y_{\rm total} + \frac{1}{\rg-1} Y_{\rm decay},
\end{equation}

\noindent where the direct photon excess, \rg $\equiv N_{\rm total}/N_{\rm decay}$, is known from measurements of the direct photon cross section~\cite{ppg042,ppg060}.  $Y_{\rm decay}$ is estimated from \piz and $\eta$-hadron correlations which compose more than 95\% of the decay background.   The daughter photon correlations are determined from those of their parent mesons by a simulation which takes into account decay kinematics as well as detector resolution, acceptance and efficiency~\cite{ppg090}.

\begin{figure}[h]
\begin{minipage}{18pc}
\includegraphics[width=19pc]{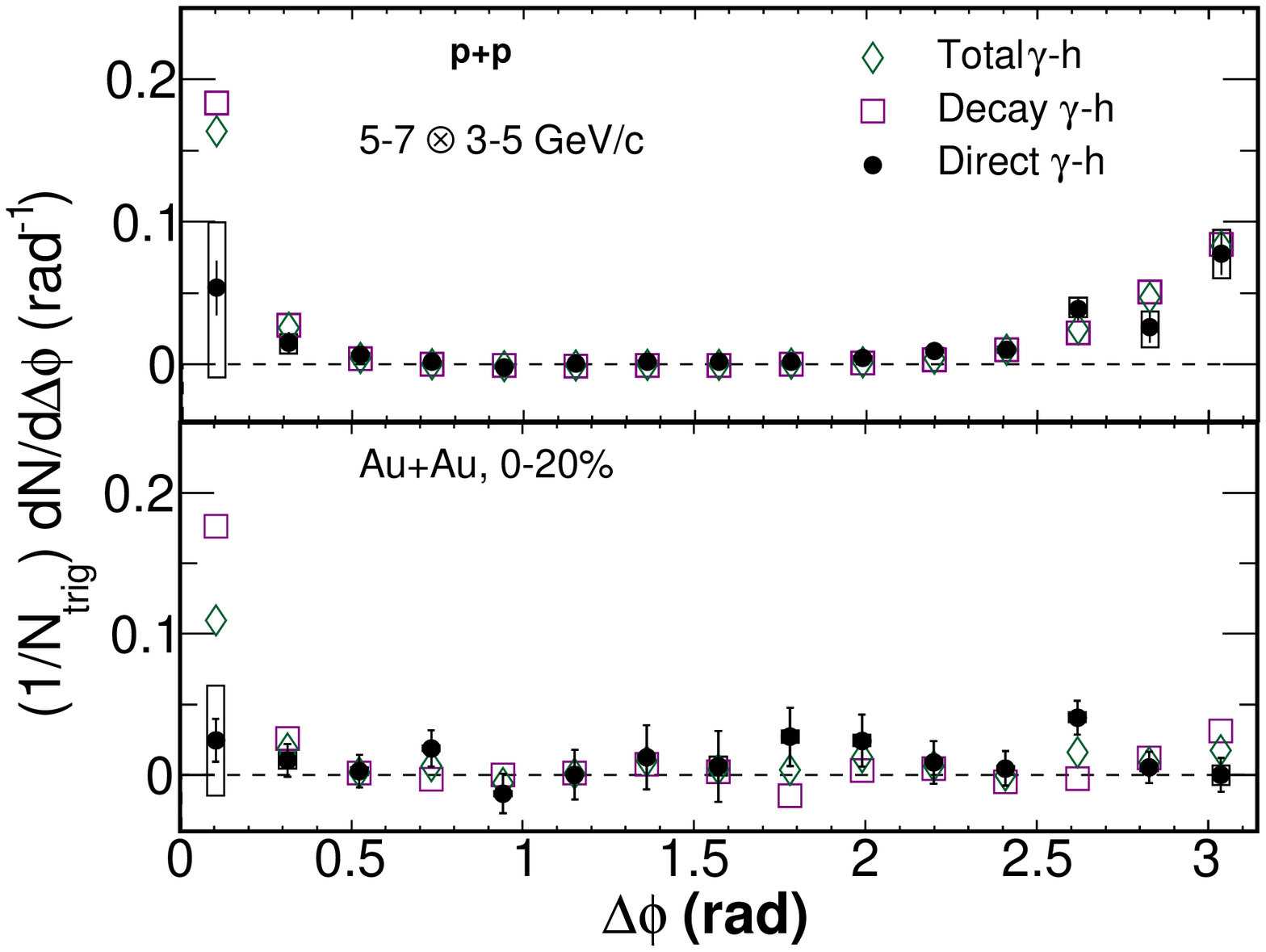}
\caption{\label{fig:dphi_stat} Azimuthal correlations of charged hadrons with inclusive, decay and direct photons in $p$+$p$ and central Au+Au~\cite{ppg090}.}
\end{minipage}\hspace{1pc}%
\begin{minipage}{19pc}
\includegraphics[width=20.5pc]{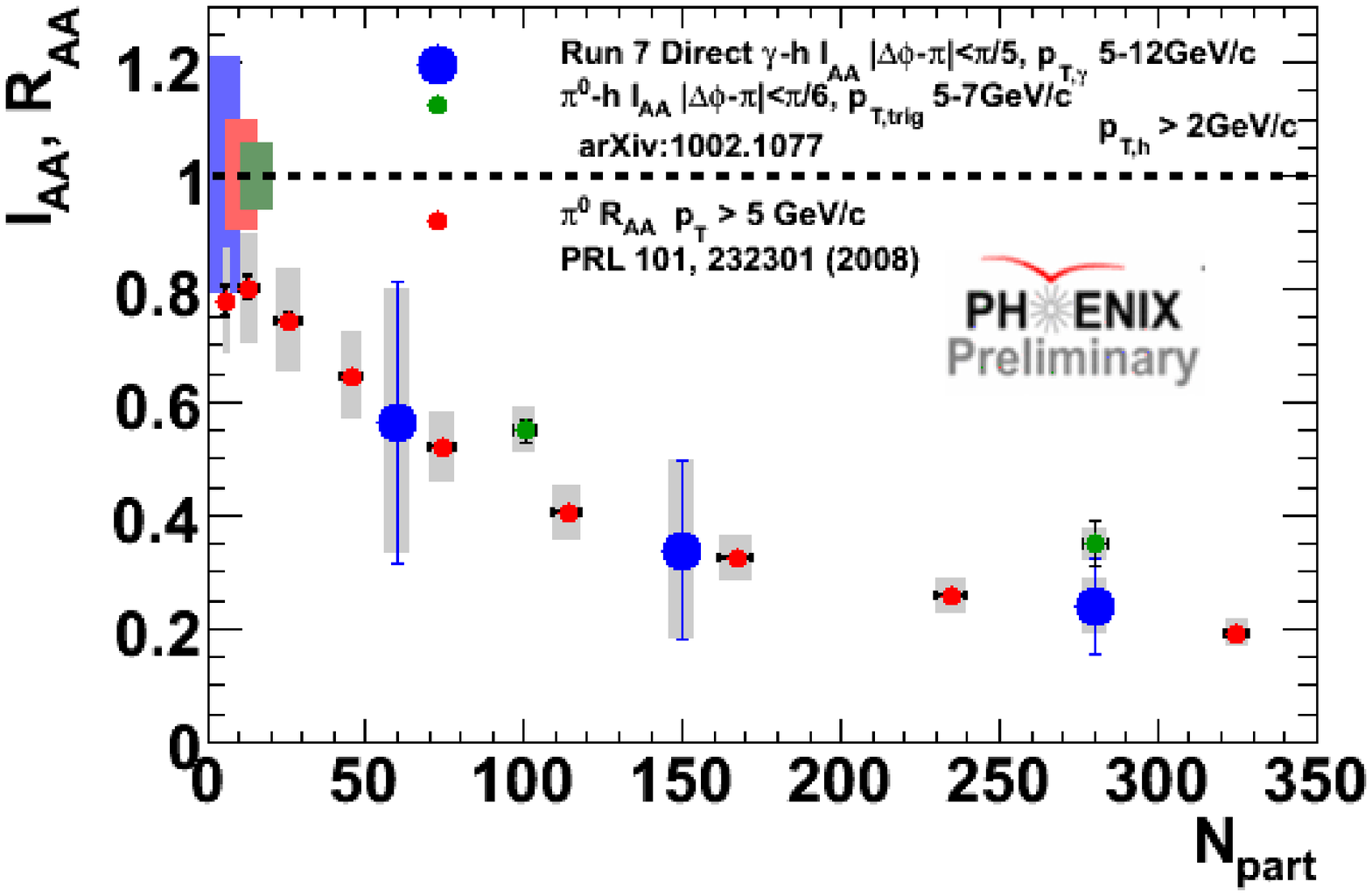}
\caption{\label{fig:iaa} 
Nuclear modification factors for \piz (\raa) and direct photon-associated and \piz-associated charged hadrons (\iaa). 
}
\end{minipage} 
\end{figure}

 Figure~\ref{fig:dphi_stat} shows examples of azimuthal correlations, $Y(\Delta \phi)$, for direct photons in $p$+$p$ and central Au+Au collisions, as well as the corresponding total and decay photon correlations used in the statistical subtraction~\cite{ppg090}.   On the near-side $Y_{\rm direct}$ is consistent with zero, as would be expected at LO.  The disappearance of back-to-back correlations in central Au+Au, a phenomenon well-known from di-hadron correlations, is also evident in $Y_{\rm direct}$.   Fig~\ref{fig:iaa} shows the nuclear modification factor \iaa, which is the ratio of $Y$ in Au+Au to $Y$ in $p$+$p$, as a function of collision centrality as quantified by the number of participants.   Also shown is \iaa for \piz-hadron correlations~\cite{ppg106} as well as \raa for single \piz's~\cite{ppg080} which both demonstrate a level of modification consistent with the direct photon correlations within sizable uncertainties.   The parton path-length through the medium is expected to vary between the three observables such that for a medium with an extended region of partial transmission, one should observe a different level of modification amongst these observables~\cite{zoww}.   In contrast, the results suggest that the medium is sufficiently absorbent that the path-length difference is beyond the sensitivity of the current measurements.   Although initial measurements of the fragmentation function using direct photon correlations have been made (not shown), no modification is apparent within the sizable  statistical and systematic uncertainties~\cite{megan}.  

\begin{figure}[h]
\begin{minipage}{18.5pc}
\includegraphics[width=20pc]{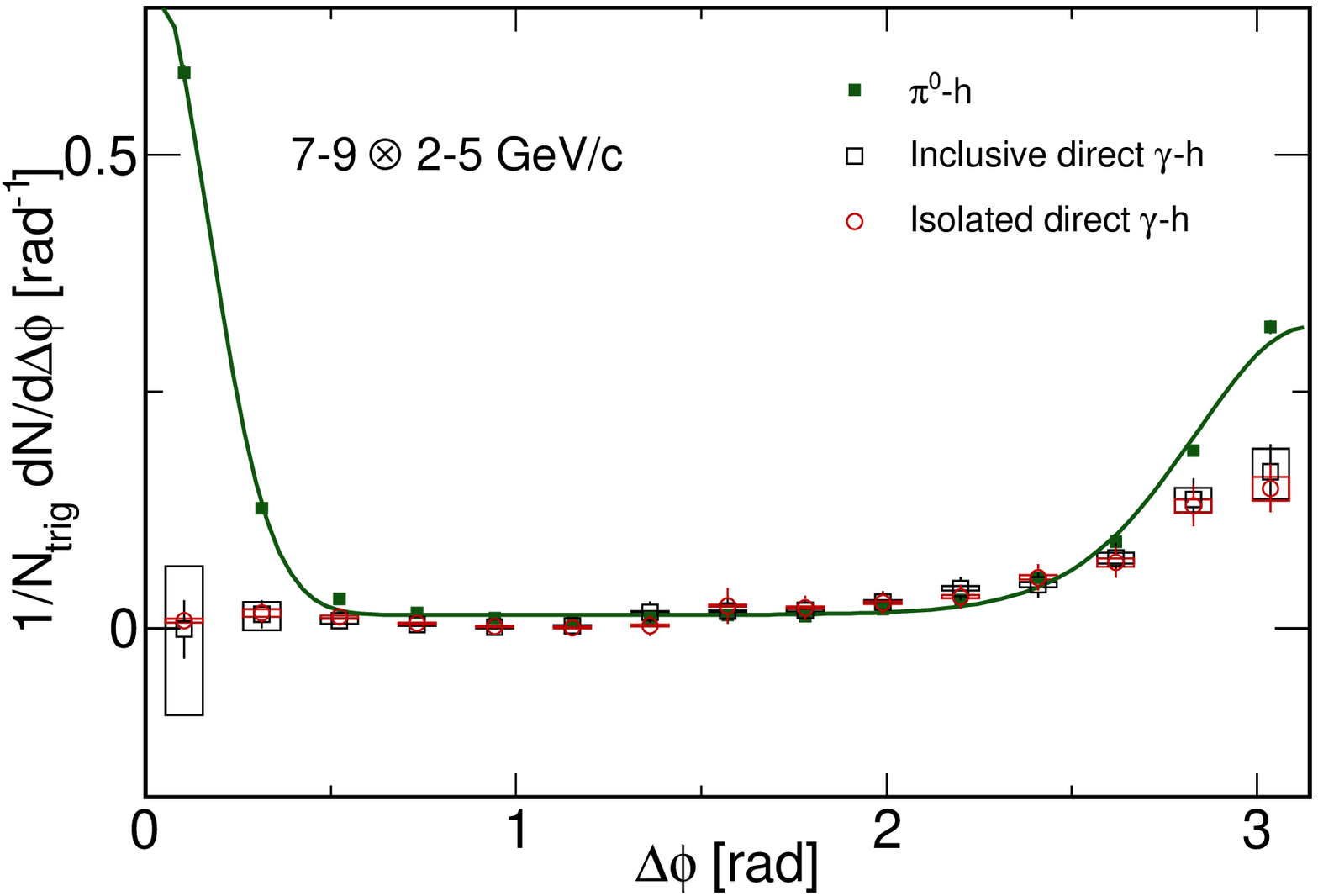}
\caption{\label{fig:dphi_iso} Azimuthal correlations of charged hadrons with \piz's, inclusive and isolated direct photons~\cite{ppg095}.}
\end{minipage}\hspace{1pc}%
\begin{minipage}{18.5pc}
\includegraphics[width=20pc]{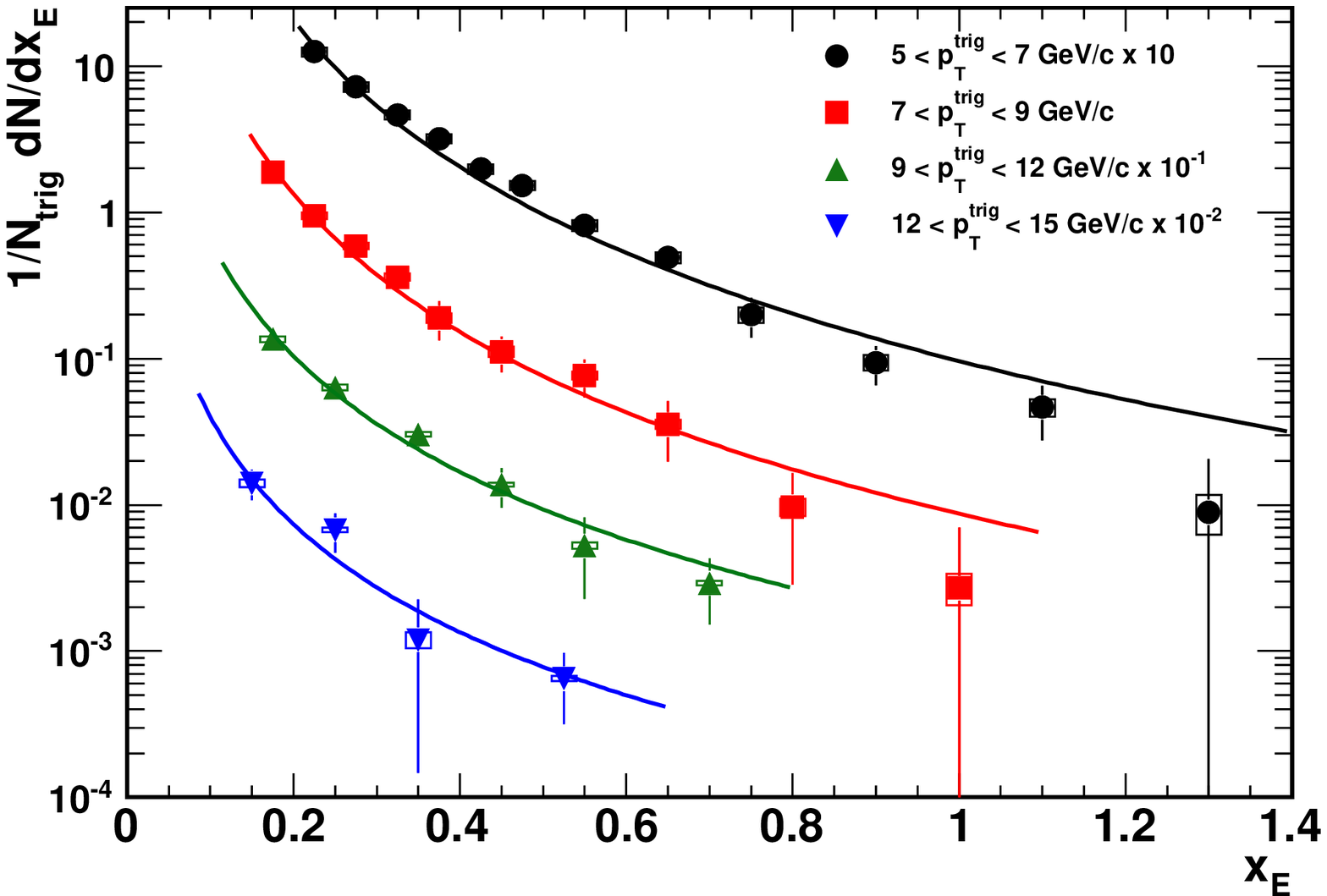}
\caption{\label{fig:xe} Charged hadron yield per isolated direct photon as a function of \xe (scaled for visibility) with power-law fits~\cite{ppg095}.}
\end{minipage} 
\end{figure}

The uncertainties in the direct photon correlations are predominantly due to the statistical subtraction of the large decay background.   Higher precision may be achieved by removing decay photons event-by-event.  This is accomplished by tagging decay photon pairs from \piz and $\eta$ based on their invariant mass and by applying an isolation cut.   Given a finite detector acceptance and efficiency, a residual background remains after the application of these cuts which is subtracted at the statistical level~\cite{ppg095}.  These methods are difficult to apply in high multiplicity nuclear collisions, but are relatively straight-forward to apply in $p$+$p$.  Figure~\ref{fig:dphi_iso} compares azimuthal correlations for direct photons using the tagging and isolation to those obtained from a fully statistical subtraction.   The uncertainties are notably reduced.  To study the fragmentation of the away-side jet, the quantity \xe ($\equiv \vec{p}_{\rm T,\gamma} \cdot \vec{p}_{\rm T, h}/|p_{\rm T,\gamma}|^2$) is used as a proxy for the fragmentation variable $z$, which is a good approximation for $p_{T,\gamma} \approx p_{jet}$. Figure~\ref{fig:xe} shows the away-side per-trigger yield as a function of \xe for isolated direct photon triggers in $p$+$p$ collisions.

\begin{figure}[h]
\begin{minipage}{18.5pc}
\includegraphics[width=19.5pc]{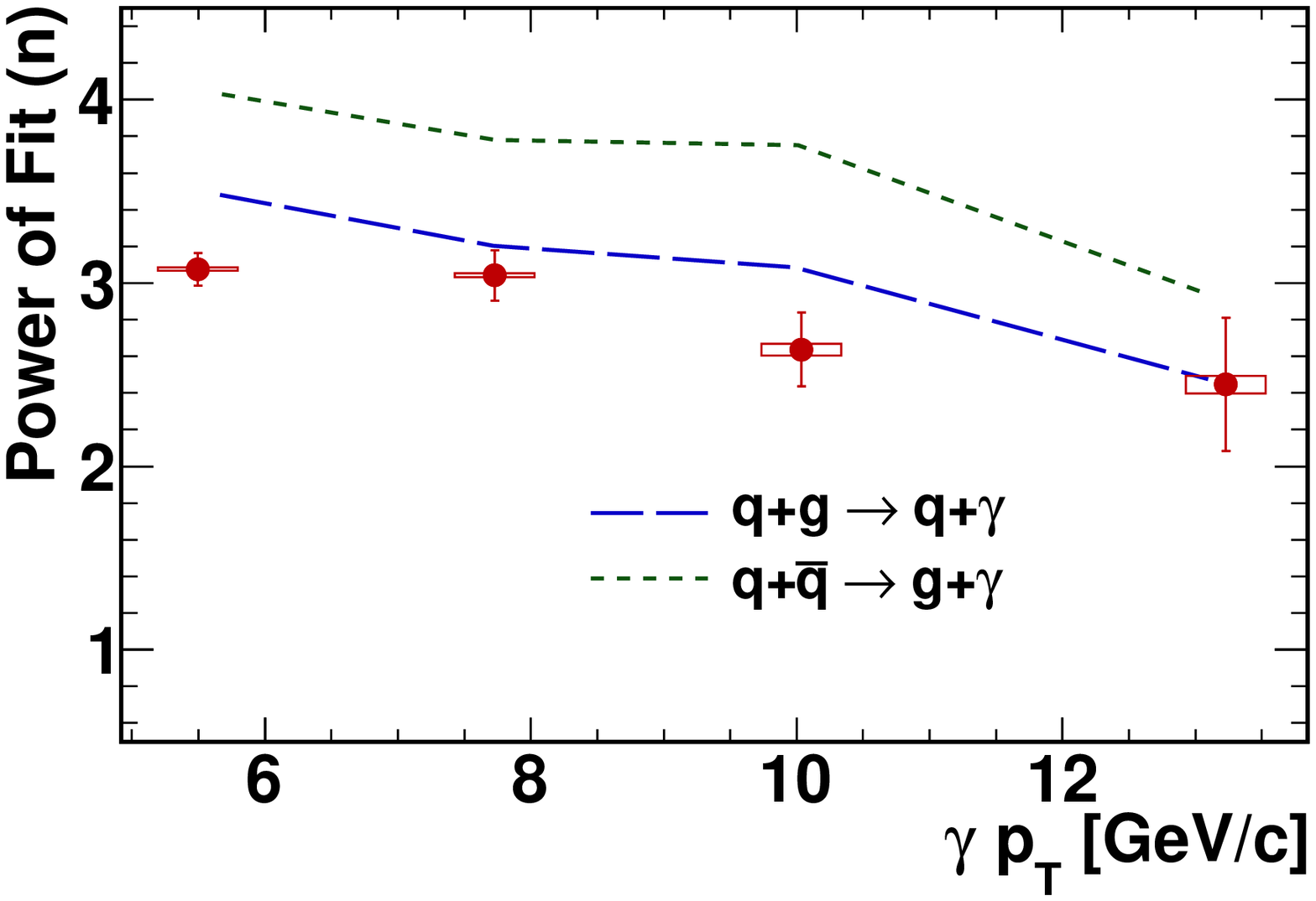}
\caption{\label{fig:slope}Power parameter $n$ of fits to the \xe distributions of direct photon triggers~\cite{ppg095}.}
\end{minipage}\hspace{1pc}%
\begin{minipage}{18.5pc}
\includegraphics[width=19.5pc]{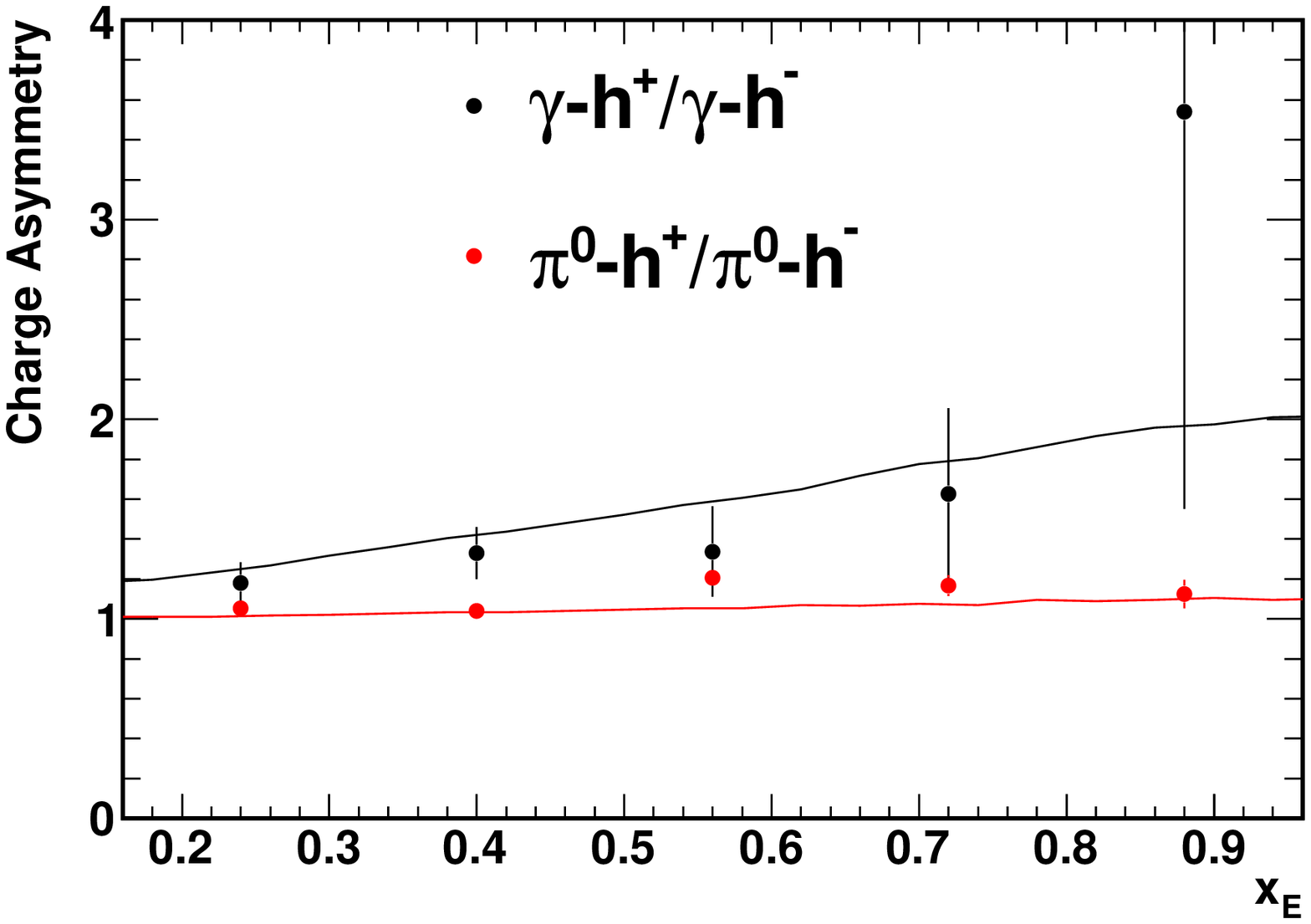}
\caption{\label{fig:charge} Charge asymmetry of partner hadrons for direct photon and \piz triggers~\cite{ppg095}.}
\end{minipage} 
\end{figure}

The values of the parameter $n$ of power-law fits to the \xe distributions (dN/d\xe $\propto \xe^{-n}$) are shown in Fig.~\ref{fig:slope}.  To test the sensitivity of the data to the underlying fragmentation function, a LO pQCD calculation was performed adding a phenomenological Gaussian \kt smearing whose width was determined from data~\cite{ppg095}.   The calculation was performed for both the Compton scattering and annihilation sub-processes, which correspond to quark and gluon fragmentation, respectively.  The data lie closer to the calculation for the Compton sub-process which is expected to dominate.   Photons are more likely to oppose an up than a down quark due to the valence quark content of the proton and the dependence of the Compton scattering amplitude on the electric charge of the quark.  Hence one should observe an excess of positive charge for away-side hadrons which is confirmed by data in Fig~\ref{fig:charge}.   Also shown is the LO+\kt smearing calculation.

\clearpage

\section{Conclusions}

We have presented results on two-particle correlations using direct photon triggers in $p$+$p$ and Au+Au collisions.  
Using a purely statistical method, we bserve the disappearance of back-to-back correlations in central Au+Au collisions similar to previous observations in the di-jet channel.  The level of suppression is consistent to that observed for single and di-hadrons, within the current statistical and systematic errors.  In $p$+$p$ collisions, more precise results were obtained by the event-by-event identification of a large fraction of the decay photon background contribution.  This was achieved using invariant mass tagging of decay pairs as well as isolation cuts.  The data were interpreted in terms of a model in which a Gaussian \kt-smearing was applied to LO pQCD.  The recoil jet was found to be compatible with quark fragmentation rather gluon fragmentation, as expect for sample dominated by LO quark-gluon Compton scattering.  A charge asymmetry of the away-side jet fragments was observed, further confirming the dominance of the quark-like nature of the away-side jet fragmentation.  The application of event-by-event methods points the way to future studies in which the uncertainties on the heavy-ion measurement may be reduced.  Furthermore, the results demonstrate that direct photons are an effective quark jet tag, a feature which may be exploited in heavy-ion collisions to study the color factor dependence of parton energy loss.

\end{document}